\documentstyle[11pt,newpasp,twoside]{article}
\newcommand{\gal}{NGC~4258}
\newcommand{\vel}{~km~s$^{-1}$}
\newcommand{\about}{$\sim$}
\newcommand{\freq}{22~GHz }

\def\edcomment#1{\iffalse\marginpar{\raggedright\sl#1\/}\else\relax\fi}
\reversemarginpar

\begin{document}
\title{Recent Progress on a New Distance to \gal}
\author{E.M.L. Humphreys, A.L. Argon, L.J. Greenhill, 
J.M.  Moran, M.J. Reid}
\affil{Harvard-Smithsonian Center for Astrophysics, 60 Garden Street,
       Cambridge, MA 02138}
 
\begin{abstract}
We report on our ongoing, high-resolution
study of H$_{2}$O masers in the innermost parsec of \gal. 
Over thirty epochs
of VLBA and VLA data, taken over six years, are being used to
monitor the velocities, accelerations, positions and proper
motions of water masers rotating in a warped, Keplerian disk
about a supermassive central object. Our extensive monitoring
results in an improved accuracy in the distance determination
to this galaxy (i) via a reduction in experimental
random errors due to a longer time-baseline than in previous
work; and (ii) via a better modeling of sources of systematic 
error, such as disk eccentricity. These data can therefore yield
an extremely accurate geometric distance to \gal.
\end{abstract}

\section{Introduction}

\gal\,\,is a nearby (\about7~Mpc) Seyfert~2/LINER that has
been extensively studied in recent years due to a sub-parsec, 
circumnuclear disk revealed by VLBA water maser observations 
at \freq (Miyoshi at al. 1995; Herrnstein et al. 1999, H99 hereafter). The 
current picture of the maser disk stems from the contributions 
of several groups. Notably, Nakai et al.\,(1993) discovered the 
existence of high-velocity features in the \freq spectrum, at 
v$_{sys}\pm$1000\vel, providing an indication that the maser 
emission could be originating from a rotating disk. 
Subsequent 
single-dish monitoring programs measured widespread velocity drifts 
in the systemic emission of \about9\vel~yr$^{-1}$, whereas negligible 
drifts were seen for the high-velocity spectral outliers 
(Haschick et al.\,1994; Greenhill et al.\,1995a; Nakai et al.\,1995). 
These data were found to be consistent with the 
line-of-sight centripetal accelerations expected for masers located in a 
disk, rotating about a supermassive black hole ($\sim$4 10$^7$ M$_{\sun}$) 
at \about7~Mpc. 
In this interpretation (Watson \& Wallin\,1994; Greenhill et al.\,1995b; 
Moran et al.\,1995), the high-velocity 
maser features should be very close to, or located on, the mid-line of 
the disk (the diameter where the sky plane intersects the disk) and the 
systemic masers are located on the front side. The existence of a 
disk of sub-pc extent was confirmed by Miyoshi et al.\,(1995) using the VLBA in observations 
which showed that the 
rotation curve of the high-velocity masers is very close to Keplerian 
(to $<$1~\%).  Multi-epoch VLBA observations led to the 
measurement of proper motions for the systemic features of 
\about30~$\mu$as~yr$^{-1}$ (H99), data consistent with the 
interpretation that the maser gas is moving ballistically in the rotating 
disk of \gal.

\begin{figure}
\plotfiddle{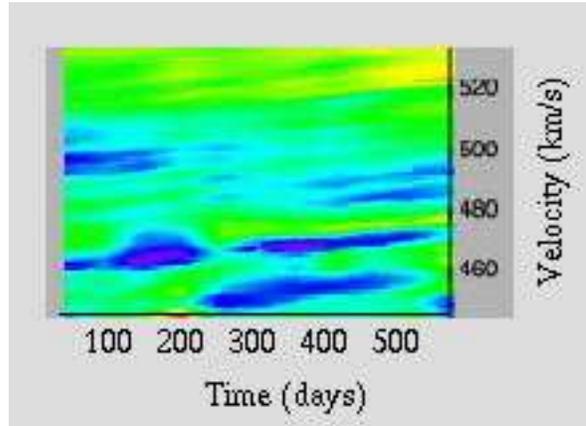}{5.5cm}{0}{100.0}{100.0}{-120}{0}
\caption{Flux density as a function of Doppler velocity
and time for a  portion of the systemic maser spectrum measured at 12
epochs between 1998-2000. The maximum flux density is around 6 Jy, 
represented by the darkest grayscale in this plot. 
\label{fig1}}
\end{figure}

\begin{figure}[t]
\plotfiddle{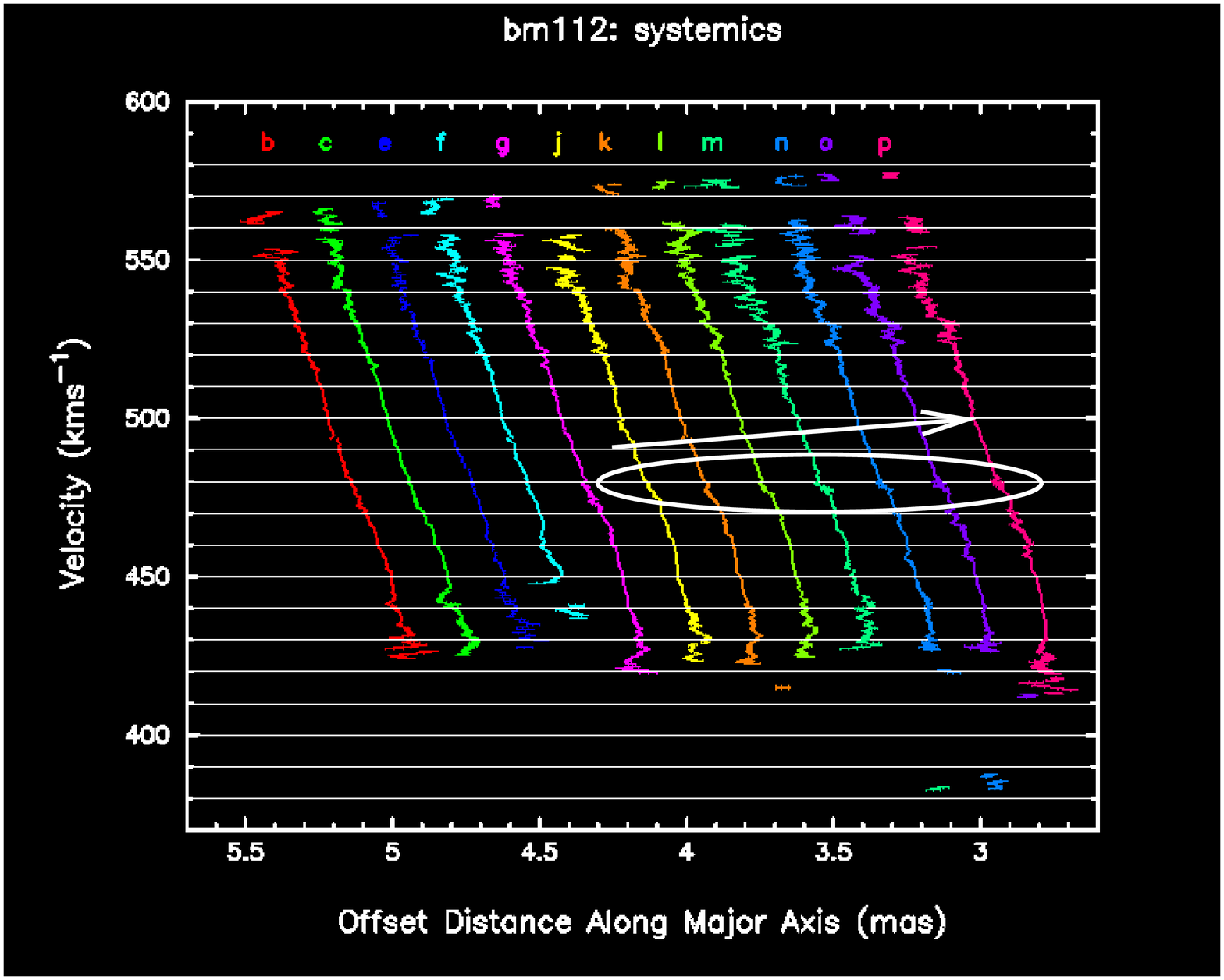}{6cm}{0}{30.0}{30.0}{-120}{0}
\caption{Rotation curve for systemic maser emission at each epoch
of observation. Different epochs have been offset along the x-axis
for clarity.
\label{fig2}}
\end{figure}

\begin{figure}[t]
\plotfiddle{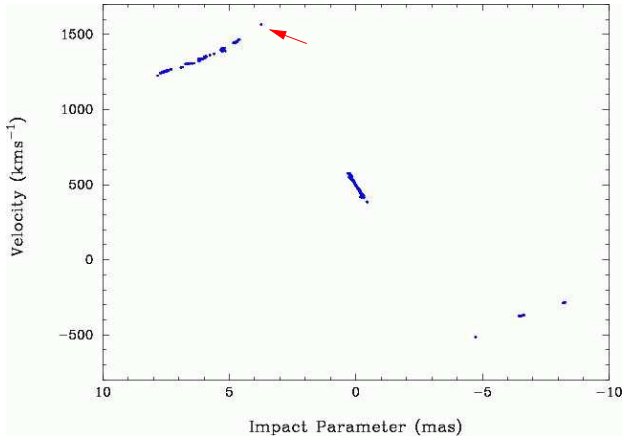}{5.5cm}{0}{40.0}{40.0}{-130}{0}
\caption{Rotation curve for the 12 epochs of data overlaid, including
the high-velocity features. The arrow points to the newly-detected, highest
velocity feature at 1562 \vel. 
\label{fig3}}
\end{figure}

\begin{figure}
\plotfiddle{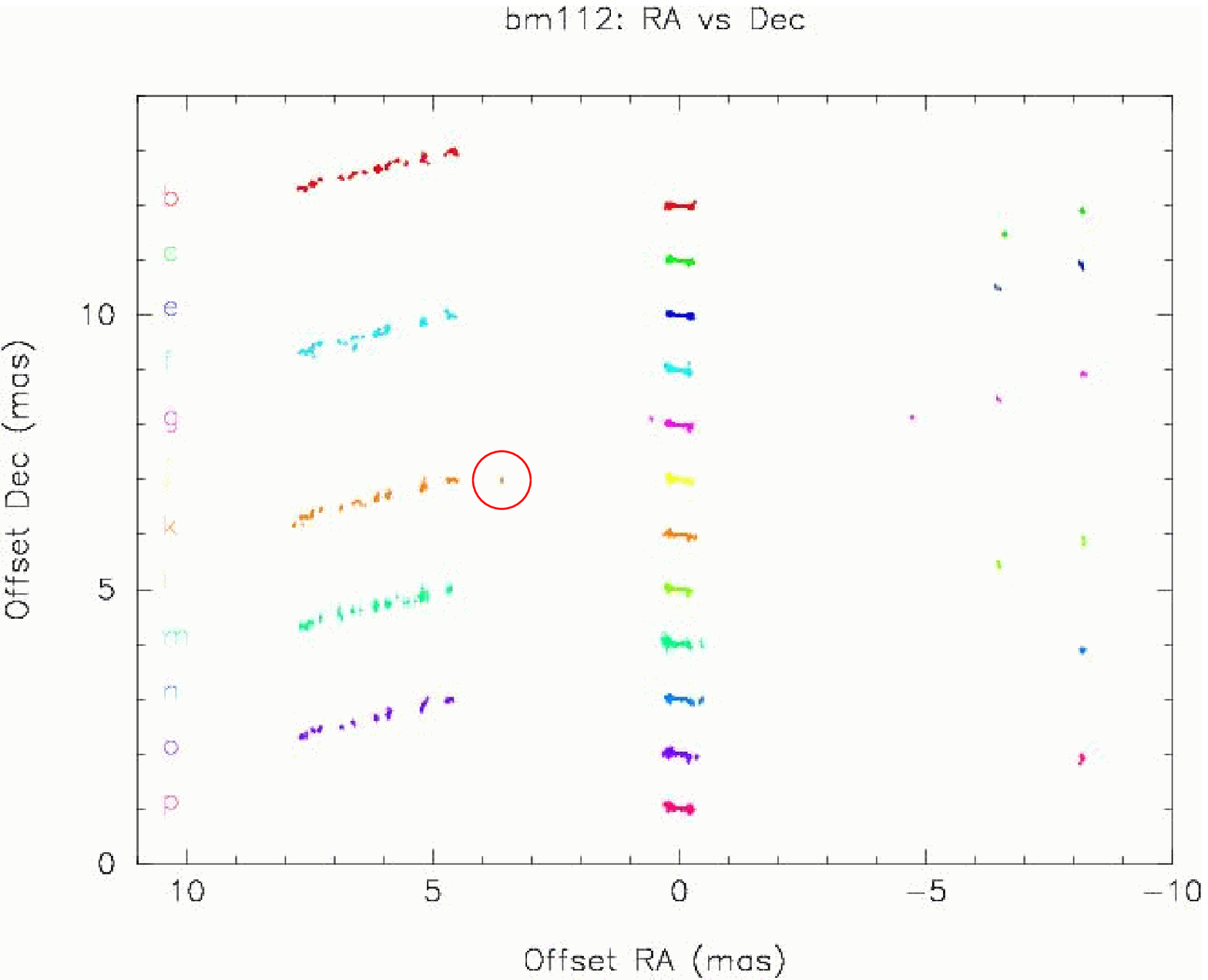}{6.0cm}{0}{50.0}{50.0}{-130}{0}
\caption{Sky  positions for the maser emission at each epoch. 
Note that the redshifted features were observed for 5 of the 12 epochs; 
the blueshifted features were observed for 7 epochs. The circle
marks the position of the highest velocity feature at 1562 \vel.
\label{fig3}}
\end{figure}

\section{This Work: A New Distance}

H99 measured a ``geometric'' distance of
7.2$\pm$0.5 Mpc to \gal. The 7~\% error in this quantity
includes components from both random and systematic errors, 
with the systematic error dominating over the random component
by a factor of 2. This distance was calculated from 4 epochs
of VLBA data taken over 3 years. Significantly, H99 
found the {\it same answer  for the distance to within $\sim$1\% using 
two broadly independent methods}, one using the maser feature accelerations 
and the other using the proper motion data.

With such an accurate distance to \gal, an obvious
question is why pursue this number to an even greater
accuracy? The answer is the role that the geometric
distance to this galaxy could play in the calibration 
of the extragalactic distance scale i.e. that \gal\,\, might
replace the still controversial distance to the LMC as the anchor for the
zero point in the Cepheid calibration. It has the potential to do this
with great accuracy, obtaining a value for H$_{0}$ which is accurate 
to 5\% or better, if  we can drive down the total error in the 
geometric distance measurement by a factor of 2 - 3.  To date, we 
have been working on reducing the  random component of the error 
and will address the reduction of systematic errors in a later paper. 
We  can achieve this (i) by increasing the number of epochs 
and (ii) by increasing the time baselines in our dataset 
(since $\sigma_{random}$  $\propto$ $N_{epochs}^{-1/2}(N_{epochs} \Delta T)^{-1}$).  
So far, we have 
increased the number of epochs to  17 imaging epochs that resolve the
disk structure and 36 epochs that provide spectra 
(17 VLBA, 14 VLA, 5 Effelsberg) over 
a duration of 6 years.

\section{New VLBA Additions to the Dataset}

In this section we highlight the most recent additions to our archive, 
12 epochs of VLBA data taken over 2 years (1998-2000). Figure 1 shows 
a portion of  
the systemic maser spectra from these data. 
It is clear from this Figure 
that some systemic maser features persist throughout the two years, 
and also that they undergo a drift in Doppler velocity. The velocity-position 
diagram for systemic masers is shown in Figure 2. In Figure 2, 
 it is evident that substructure in the disk
exists, that is persists between epochs, and that it changes velocity
consistent with the centripetal acceleration of the disk.
We believe that this
provides further evidence that we are tracking `persistent features' moving 
in the disk. Figure 3 goes on to show the rotation curve for the 12 epochs 
plotted overlaid, the arrow pointing to the new detection of masing material 
at the highest velocity that has ever been observed, at 1562 \vel. 
 This feature, whose sky position is marked by the circle in Figure 4, is 
now the closest high-velocity maser feature to the center of the disk. 
 This
means that the inner edge of the high velocity features quite
closely corresponds to the disk radius of the systemic features, but does not
lie inside this radius (cf H99).

{\small
\begin{table}
\caption{Current best-fit parameters for maser disk ($\chi^2$=1.8), see text
for label definitions.}
\begin{tabular}{lllllllll}
\\
\hline
v$_{sys}$& x$_0$&y$_0$& M/D &$\alpha_0$&$\alpha_1$&$\alpha_2$ &$i_0$ & $i_1$ \\

(kms$^{-1}$)&(mas)&(mas) &(10$^7$$M_{\sun}$&(deg)&(deg& (deg&
(deg)  & (deg \\
     &    &   &Mpc$^{-1}$)&&mas$^{-1}$)& mas$^{-2}$)&
  &mas$^{-1}$) \\
\hline
\\
474.1 &$-$0.132 & 0.548 & 0.526  & 65.3 &2.3 &$-$0.23  &71.6 &2.4 \\
\\
\hline
\end{tabular}
\end{table}

}

\section{From Data to Distance}

In order to calculate the distance, we require maser feature accelerations 
or/and proper motions and the polar coordinates of maser features in the 
disk. 
We describe how the necessary quantities are obtained below.

{\bf (i) Accelerations \& Proper Motions:}
The method used here to measure feature accelerations and proper motions 
differs from that used by H99. Since the time intervals between
epochs were relatively long in H99, and the number of epochs limited to 4, 
individual features in the spectra or images could not be tracked `by eye'.
 Instead a 
Bayesian analysis technique was employed in order to match up maser features between
epochs, resulting in measured accelerations and proper motions. 
Since we have a more frequent time sampling, over a longer time baseline, 
we are able to track unambiguously  spectral features between epochs. 
This is achieved in practice by using
 a non-linear, Gaussian fitting routine which 
decomposes {\it simultaneously} the spectra of up to 36 epochs of data 
into individual Gaussians, each corresponding to a maser `feature'. The
routine fits for feature Doppler velocity (which can be used to
determine  accurate sky
positions for the proper motion measurements) and amplitude. It also
fits for feature  line-of-sight acceleration, provided by the
drift in Doppler velocities of each feature among epochs, and the
time derivative of this quantity, i.e the jerk.

{\bf (ii) Maser Disk Coordinates:}
The deprojection of maser features from the sky plane requires
the construction of a 3D disk model that introduces the disk-modeling 
systematic errors to the distance measurement. 
In order to deproject the maser sky positions we employ a 9-parameter 
$\chi^{2}$ fitting routine which fits for the systemic, galactic velocity (v$_{sys}$), 
dynamical disk center (x$_{0}$,y$_{0}$), black hole mass/distance (\mbox{${\cal M}$}=M/D),
position angle warping 
($\alpha$($\theta_r$) = $\alpha_{0}$ + $\alpha_{1}\theta_r$ + $\alpha_{2}\theta_r^2$; $\alpha$ measured East of North)  and inclination angle warping (i($\theta_r$) = i$_{0}$ + i$_{1}\theta_r$) where $\theta_r$ is the radial distance
in the disk in angular units.
A ``genetic'' algorithm is employed in order to find the global minimum for 
this fit, in which the VLBA data from all 12 epochs is included. Table 1 shows 
that the best fit to date indicates that {\it the maser disk is 
significantly warped} (H99 and references therein; 
Herrnstein et al. 2003). This process yields ($\theta_{r}$,$\phi$; $\phi$
measured from the disk midline), the
angular polar disk coordinates for each maser feature, for the fit that has resulted in the smallest value of $\chi^{2}$.

\smallskip

We now have all the quantities required to calculate the distance since,
using centripetal acceleration, the distance is given by a weighted mean of
$D = G$\mbox{$\cal M$} $ \cos(\phi)\sin(i(\theta_r)) / (\theta^{2}_{r} 
\dot{v}_{los})$ for maser features, where G is the gravitational constant.
{\bf In conclusion, a preliminary analysis indicates that so far we have 
succeeded 
in reducing the 
random component of the distance error to a third of 
its previous value. The total error (random + systematic) on 
the distance measurement to NGC 4258 has now been reduced from 7\% to 5.5\%.}


\begin{references}
{\small
\reference Greenhill, L. J., Henkel, C., Becker, R., et al., 1995a, \aap, 304, 21

\reference Greenhill, L. J.,  
            Jiang, D. R., Moran, J. M., Reid, M. J., et al., 1995b, 
            \apj, 440, 619 
\reference Haschick, A. D., Baan, W. A., 
         Peng, E. W. 1994, \apj, 437, L35
\reference Herrnstein, J. R.,  
           Moran, J. M., Greenhill, L. J., et al., 1999, Nat., 400, 539 (H99)
\reference Herrnstein, J. R.,  
           Moran, J. M., Greenhill, L. J., Trotter A. S., 2003, in prep
\reference Miyoshi, M., Moran, J. M., Herrnstein, J.
         et al., 1995, Nat., 373, 127 
\reference Moran, J. M., Greenhill, L. J., 
           Herrnstein, J., et al., 1995, PNAS, 921, 1427 

\reference Nakai, N., Inoue, M., Miyoshi, M., 
        1993, Nat., 361, 45

\reference Nakai, N., Inoue, M., Miyazawa, K., 
        et al., 1995, PASJ, 47, 771

\reference Watson, W.D., Wallin, B.K. 1994, 
            \apj, 432, L35
}
\end{references}
\end{document}